\documentclass[reprint,aps,groupedaaddress]{revtex4-2}
\usepackage{times}
\usepackage{graphicx}
\usepackage{amsmath}
\usepackage{amssymb}
\usepackage{graphicx}
\usepackage{lmodern}
\usepackage{wrapfig}
\usepackage{xcolor}
\usepackage{hyperref}
\hypersetup{
    colorlinks = true,
    linkcolor = blue,
    anchorcolor = blue,
    citecolor = magenta,
    filecolor = blue,
    urlcolor = blue
    }
\usepackage{mathrsfs}
\usepackage{color}
\usepackage{csquotes}
\MakeOuterQuote{"}
\usepackage{gensymb}

\begin{document}

\title{Activity-Induced Annealing leads to Ductile-to-Brittle Transition in Amorphous Solids}

\author{Rishabh Sharma}
\author{Smarajit Karmakar}
\email{smarajit@tifrh.res.in}
\affiliation{
Tata Institute of Fundamental Research, 
36/P, Gopanpally Village, Serilingampally Mandal,Ranga Reddy District, 
Hyderabad, 500046, Telangana, India } 

\begin{abstract}
Investigating the behavior of amorphous solids under various external loading conditions continues to be an intriguing area of research with significant practical implications. In this study, we demonstrate the utilization of self-motility as a means to anneal glasses and use that as a means to fine-tune the failure mode of the system under uniaxial tensile deformation. We begin by highlighting the annealing effects of activity and draw parallels with other well-known mechanical annealing processes, such as oscillatory shearing (both uni- and multi-directional). Furthermore, we explore the annealing effects in the presence of open boundaries, observing enhanced surface relaxations due to activity. By implementing various activity-induced annealing protocols, we successfully induce a transition in the failure mode from ductile to brittle. This is demonstrated via performing tensile tests on the glass samples resulting from the active-annealing process. The intricate effects of geometry on the formation of shear bands are also examined. We find that samples having an aspect ratio greater than one fail via shear band formation, owing to their formation angle of $45\degree$ from the strain axis. In conclusion, we introduce a novel method for producing well-annealed glasses in silico and establish a correspondence between sheared and actively driven glasses.
\end{abstract}

\maketitle

\section{Introduction} 
Glasses are out-of-equilibrium materials formed when a supercooled liquid is cooled below its (empirically defined) glass transition temperature ($T_G$), at which point the liquid viscosity exceeds $10^{12}$ Pa.s \citep{biroli_aging, BiroliRMP, KDSAnnualReview, KDSROPP,Bertheir_Ediger, Cavagna}. This enormous change in viscosity occurs over a small temperature change, usually just a few tens of kelvins. The central problem of glass transition is understanding how such a large dynamical change can occur in a system without any significant structural change. Additionally, when the temperature is kept below $T_G$, the relaxation time becomes so large that, for all practical purposes, one observes an amorphous solid. The mechanical properties of such solids are a topic of significant research interest due to their importance in various practical applications and in future material design. Because of the inherent non-equilibrium nature of glass formation, the material properties depend not only on their composition but also on the preparation protocols. Owing to the lack of long-range order, glasses do not suffer from defects like grain boundaries, which are present in their crystalline counterparts and can lead to unwanted structural and optical features. Constant efforts to alter or tune the material's mechanical properties as desired have led to the establishment of various annealing methods. Such methods aim at achieving better material properties, with the recent discovery of ultra-stable glasses being a notable development \cite{ediger_etal}.

The mechanical failure mode of glasses can be tuned in various ways. For instance, confining geometry can change how glasses break. In recent experimental and simulation works \citep{Luo_etal, Volkert_etal, paul_etal}, it has been found that amorphous solids can show very different mechanical failure behavior under uniaxial tensile deformation at bulk and nano-scale. For example, a material that can withstand only 2\% strain at bulk before complete brittle-like failure can be deformed up to 200\% if the dimension of the sample is at a few hundred nano-meter scales \cite{Luo_etal}, with neck-like failure. The thickness of the necking region can become as small as a few atomic diameters just before the total failure. This drastic change in failure mode from brittle to ductile with changing sample size when going from bulk to nano-scale highlights the importance of surface relaxation in the failure process. In \cite{paul_etal}, the existence of a critical aspect ratio (ratio of height to width in a two-dimensional sample or ratio of length to cross-sectional linear dimension for a three-dimensional sample) is observed. Below this critical aspect ratio, the material shows neck-like ductile failure, which then crosses over to cavity-dominated brittle-like failure for aspect ratios larger than the critical value.

Similarly, one can modify the interaction range of the constituent particles to transition from brittle (smaller range of inter-particle interaction) to ductile (larger range of inter-particle interaction) failure \cite{dauchot_etal}. Introducing impurities or inclusions having different mechanical properties than the embedding matrix can also lead to changes in failure mode\cite{bhowmik_etal}. For example, one could cross over from a heterogeneous shear band-mediated yielding behavior to homogeneous ductile yielding with increasing appropriate inclusions, as is done in various micro-alloying methods. The degree of annealing or changing the cooling rate during preparation has a huge impact on the resulting glass' mechanical properties. For example, a glass that is prepared through a rather high cooling rate will be poorly annealed. If such a glass is subjected to simple shear, it shows a more ductile yield. In contrast, a glass that is prepared at a slow cooling rate or produced via other methods that can anneal the solids much better will generally be more well-packed and have much larger elastic moduli. Unfortunately, such a solid will also show catastrophic brittle failure via shear band formation when subjected to deformations \cite{ludovicPNAS}. In general, the better-annealed a glass is, the more "brittle" it tends to be.

The pursuit of creating increasingly well-annealed glasses has implications not only for exploring the mechanical properties of these systems but also for getting closer to the proposed "ideal glass." This ideal glass can be considered the most well-packed yet disordered state of matter, whose configurational entropy, much like that of a crystal, is zero. Grasping these concepts is crucial for probing the nature of the glass transition at a deeper level. Recently, significant breakthroughs have been made in developing various experimental techniques, such as Physical Vapour Deposition (PVD)\cite{ediger_etal}, and computational techniques like swap Monte Carlo \cite{Bertheir_SwapMC} and in silico vapour deposition \cite{singh_etal}, to achieve ever-lower energy states in glasses. Each of these methods comes with its advantages and limitations. Thus, an exploration of other generic methods which can be employed for better annealing a wide variety of disordered solids is certainly a compelling avenue of research. In this regard, in computer simulations, people have investigated different forms of mechanical deformations that can lead to the annealing of glasses. Oscillatory shearing is a particularly well-researched method among these techniques\citep{srikanth_oscilatory_shear,srikanth_finite_temp,vishnu_etal}.

In this paper, we introduce a novel form of mechanical annealing: annealing glasses through local internal perturbations facilitated by active particles. Active particles have motility that derives from either their internal energy reserves or from external sources \citep{ramaswamy_active,Narayan_etal,Wilson_etal,Janus}. 
Such motility in the context of glasses has been helpful in understanding many biological systems where dense cell packing and ATP-driven transport is the norm \citep{ramaswamy_soft_active,Janssen_active_glasses,Henkes_etal,manning_glassyness_cells,manning_tissue,nandi_active_glass_pnas}.
In our observations, we find that the annealing and yielding behavior of glasses subjected to active driving closely resembles that seen in glasses annealed using oscillatory shearing\cite{srikanth_oscilatory_shear}. This helps to strengthen the correspondence between actively driven and sheared glasses which has been observed before \cite{morse_etal}. We alluded to such similarities in an earlier work dealing with the formation of cavities in glasses \cite{umang_etal}. There, it was found that oscillatory shear cycles can lead to cavitation at a density that is much larger than the density at which one expects to see cavitation failure via only uniform expansion protocols. It was also shown that qualitatively similar results could be obtained if these samples were subjected to local deformation via active particles instead of oscillatory shear. In this current work, we take this correspondence further by focusing on the annealing effects of active dynamics. 
After establishing the annealing effects, we further demonstrate how this process can be leveraged to tune the failure mode from ductile to brittle under tensile loading conditions. These results suggest a deeper connection between annealing via oscillatory shear and active annealing.

The organization of this paper is as follows: Section~\ref{sec:model} provides details on the simulation and model. Section~\ref{sec:protocols} describes the protocols employed to generate the initial states, as well as the methods used for activity-induced annealing and tensile testing. In Section~\ref{sec:results}, we present our findings, and we conclude the paper in Section~\ref{sec:conclusion} with the summary and future directions.

 \section{Model and simulation details}\label{sec:model}

For our MD simulations, we used a well-known model glass former, the binary Kob-Anderson (KA) mixture \cite{kob_etal}. It involves 2 species of particles interacting  via Leonard-Jones potential (eq.\ref{lj_pot}),  with the larger A-type and smaller B-tyle present in the ratio of 80:20.

\begin{equation} \label{lj_pot}
V_{\alpha\beta}(r_{ij}) = 4\epsilon_{\alpha\beta} \left[ \left(\frac{\sigma_{\alpha\beta}}{r_{ij}} \right)^{12} - \left(\frac{\sigma_{\alpha\beta}}{r_{ij}} \right)^6 \right] + u(r_{ij})
\end{equation}

The energy and length scales are chosen such that  $\epsilon_{AA} = 1$ and $ \sigma_{AA} = 1 $ . The interactions  between the other combinations of particles are then given in terms of $\epsilon_{AA}$ as $\epsilon_{AB}=1.5\epsilon_{AA}$ and $\epsilon_{BB}=0.5\epsilon_{AA}$. Similarly for  length scales, we have $\sigma_{AB} = 0.8\sigma_{AA}$ and $\sigma_{BB} = 0.88\sigma_{AA}$. The cut-off range is taken to be $2.5\sigma_{ij}$, and the potential is made to go to zero smoothly at this point (by choosing $u(r_{ij})$ accordingly so as to make the slope continuous at the potential cutoff).

For imparting activity, we add an additional force $\vec{f_0}$ (eq.~\ref{eq:active_force}) to the smaller B-type particles in addition to the potential derived force.
\begin{equation}
\vec{f_0} = f_0(k_x\hat{x} + k_y\hat{y} + k_z\hat{z}).
\label{eq:active_force}
\end{equation}
The force is added along the eight diagonal directions, and the directions are shuffled after a persistence time $\tau_p$ (here taken to be $\tau_p = 4$). This results in run and tumble dynamics, and in literature, it is referred to as the 8-state clock model \cite{mandal_etal}. Momentum conservation is ensured by maintaining the sum of $k$'s to be zero component-wise throughout the simulations. Nos\'{e}-Hoover thermostat and barostat were used to maintain the desired temperatures and pressures.

The typical system size considered is $N = 10,000$, with additional simulations performed for $N = 32,000$, $N = 64,000$, and $N= 128,000$ particles to study systematic finite size effects. A very low temperature of $T=0.01$ is maintained throughout activity-driven annealing and during tensile testing. Annealing was done under periodic boundary conditions, and open boundaries were created for tensile testing and surface relaxation studies. All the simulations were conducted using our custom parallel MPI C codes.

\section{Annealing and tensile Testing protocols}\label{sec:protocols}
\textbf{Initial states:} The states for annealing were prepared by cooling a liquid equilibrated at a high temperature of $T = 1.0$ and high density  $ \rho = N/V = 1.2$ to a temperature of $0.01$. The cooling rates were varied from $\dot{T} = 10^{-1}$ to $\dot{T} = 10^{-6}$. These resulted in states with inherent state per particle energies ranging from $-6.910$ (poorly annealed) to $-7.035$ (well annealed). An ensemble average of $16$ independent samples are taken for each case.

\textbf{Annealing protocol:} The obtained states, having a range of inherent state energies, are then subjected to the local perturbations via active dynamics. The protocol involved imparting activity to all the B-type particles in the system and evolving such a system for $10^5$ time units (or $2\times10^7$ MD steps) under isothermal conditions at a low temperature of $T=0.01$. Thus, the system was left to age in the presence of various magnitudes of active forcing ($f_0$), and the resulting states were then used to perform tensile testing.

\begin{figure*}[t!]
  \centering
  \includegraphics[width=1.0\textwidth]{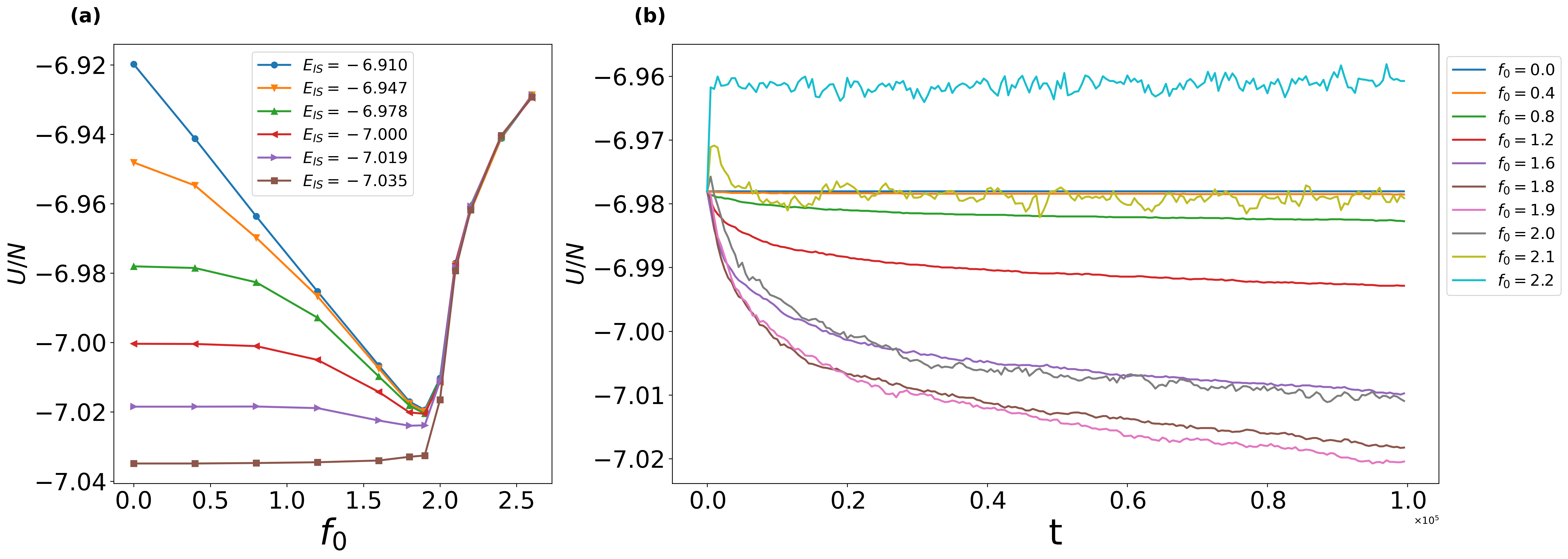}
  \caption{Annealing and yielding in glasses via local perturbations due to activity. Fig. (a) shows the effects of active dynamics (carried out for a time window of $t_w$) on glasses having various 
  inherent state energies. Up to a threshold of $f_0 = 1.9$ for this model, we see enhanced aging behavior, as seen by the system's progression towards progressively deeper minima. Beyond this threshold, the various curves collapse to a single curve, denoting a yielded state where the system no longer retains a memory of its initial preparation. A very similar transition from absorbing to yielded state is seen in oscillatory shearing, with increasing oscillation amplitude. (b) Annealing of glass having $E_{IS} = -6.947$ (corresponding to the green curve in (a)) is shown. Decaying behavior for states till $f_0 = 1.9$ can be seen, whereas, for larger forcing amplitude, we observe a yielded steady state. System size of $N = 10,000$ averaged over $16$ ensembles.}
  \label{fig:anealing}
\end{figure*}

\textbf{Energy minimization protocol:} To understand where in the energy landscape the system is during the annealing process, we sampled states at equal intervals from the isothermal aging trajectory and performed energy minimization using the well-known conjugate gradient (CG) method. The final energies plotted in Fig. \ref{fig:anealing}(a) are the average energies of the last 5 frames from the tail of the energy-minimized trajectories (Fig.~\ref{fig:anealing}(b)).  

\begin{figure}[htbp]
  \centering
  \includegraphics[width=0.5\textwidth]{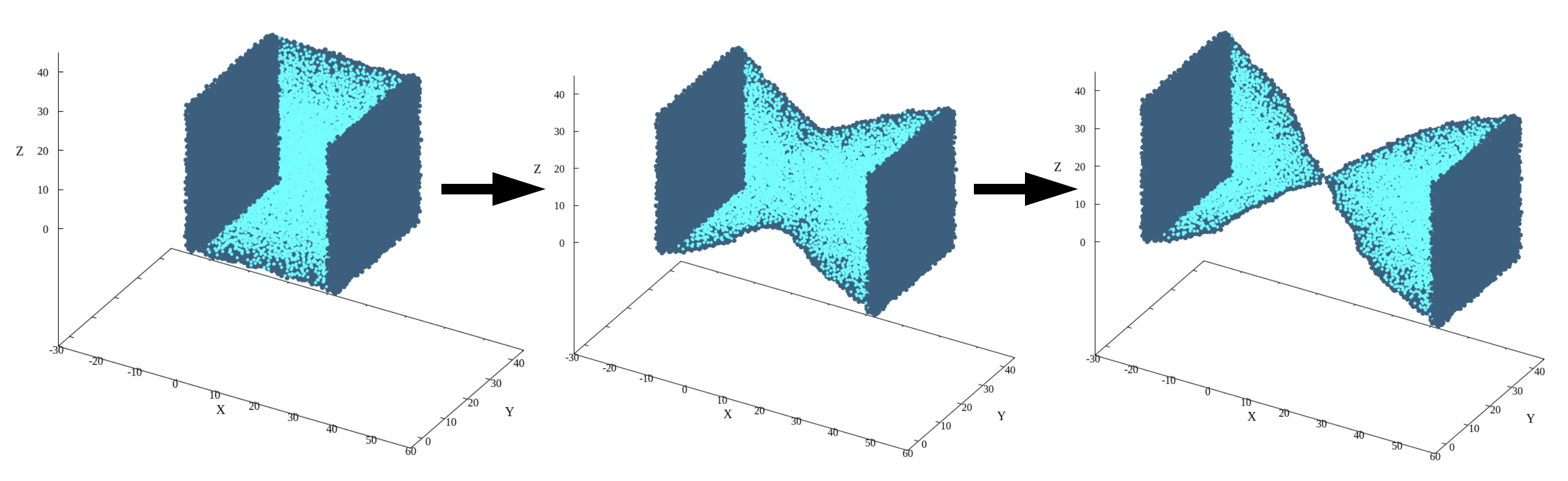}
  \caption{Schematic of the tensile testing procedure carried out in the study.}
  \label{fig:schematic}
\end{figure}
\textbf{Tensile testing protocol:} To perform tensile testing, we utilized the various annealed states and subjected them to $200,000$ Molecular Dynamics (MD) steps under zero pressure conditions. This step was crucial to prevent any pressure shock when transitioning from periodic to open boundary conditions. Subsequently, we created two walls of width $2.5\sigma_{AA}$ by freezing the particles' degrees of freedom along the two faces of the containing box (along the x - direction). Open boundary conditions were created along the other two directions. This configuration was run for an additional $200,000$ MD steps to allow the surface to "settle in" before applying a constant strain rate, denoted by $\dot{\gamma}$, to the walls in opposing and outward directions. This application induced tension in the system, and we studied the system's response to this tension. The system was pulled till complete failure. The schematic for the process is shown in Fig. \ref{fig:schematic}.

\begin{figure*}[t!]
  \centering
  \includegraphics[width=1.0\textwidth]{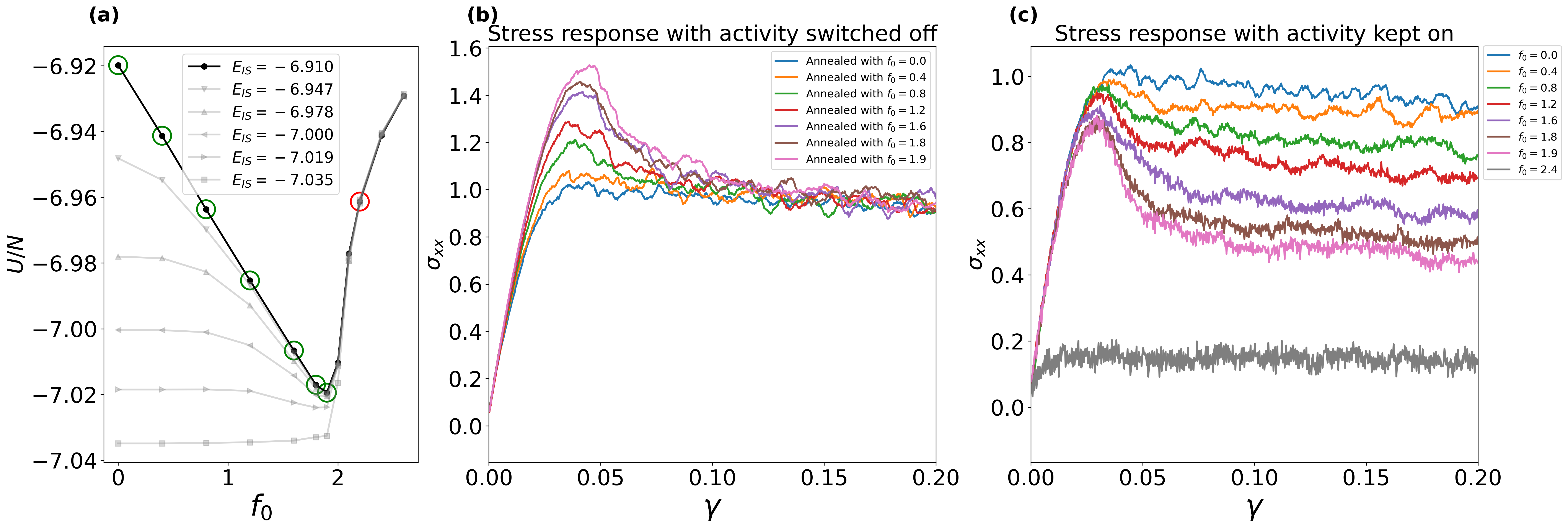}
  \caption{The states encircled on the highlighted curve in (a) are sampled for tensile testing; the green circles mark the aging (or annealing) states, and red is used for the yielded state. The reason for choosing this curve is to span the largest range of inherent energies reached through active annealing. In (b), we show the stress-strain curves obtained when these states are pulled after turning off the activity. This sequence somewhat resembles a Physical Vapour Deposition (PVD) process where highly motile particles -responsible for annealing the surface- lose their mobility after being trapped in additional layers of particles. From the stress response of these states, we see better-annealed states showing a greater stress drop, indicating an enhancement in the brittleness after an active dynamics annealing treatment. In (c), we perform the same testing, but with the annealing activity kept on. Here too, the larger stress drops are preserved. Here we also note the yielded state is unable to withstand any considerable amount of stress. These results are for a system size of $N = 10,000$ averaged over 16 ensembles and strain rate $\dot{\gamma} = 5\times10^{-5}$.}
  \label{fig:tensile_tests}
\end{figure*}

\section{Results}\label{sec:results}

Fig. \ref{fig:anealing} encapsulates the outcome of our annealing protocol. We report a trend that is remarkably similar to the one observed in \cite{10.1073/pnas.2100227118}, where the annealing is performed using oscillatory shearing and under athermal conditions. Active glasses appear to age faster, enabling them to reach lower points in their potential energy landscape than their inactive counterparts. Up to a certain threshold $f_0$, we observe annealing beyond which the energies again begin to rise and collapse to a single curve. Drawing an analogy with the yielding amplitude of oscillation ($\gamma_{max}$) seen in oscillatory shear simulations, we term this the active yielding amplitude ($f_{0}{^Y}$). In our case, we find $f_{0}{^Y} \simeq 1.9$. The various curves collapsing to a single curve beyond $f_{0}{^Y}$ signifies that the system no longer remembers its original preparation history and thus can no longer be in the aging regime. This rise in saturation energy beyond yielding amplitude is illustrated in Fig. \ref{fig:anealing}(a).
 
One also observes the degree of annealing via activity decreases as one goes to lower and lower initial state energies. Thus, poorly annealed glass responds much better to annealing via activity as compared to very well-annealed glasses. This trend, too, resonates with that reported in oscillatory shearing. One way to understand this is to realize that poorly annealed glasses have more "soft spots" or Shear Transformation Zones (STZ's) \citep{STZ_experimental,Langer,soft_spots} that are available to be triggered by the microscopic perturbations caused by active forcing. A dearth of such soft spots in well-annealed glasses also explains their low sensitivity to both active and oscillatory annealing. 

One crucial difference to note is that under athermal oscillatory annealing, the system below the yielding amplitude always reaches a limit cycle in the form of an absorbing state, and this limiting energy makes for a natural "stopping point." Such an absorbing state transition has also been observed in aging studies of topologically constrained active matter \cite{Hartmut_janssen} (again, only in the absence of temperature). In active annealing, no such absorbing states are present, partly due to the dynamics being run at a finite, albeit low, temperature. The energy below the yield point decays logarithmically, and thus we take a finite time window instead to get a sense of the limiting energies. The exact values in Fig. \ref{fig:anealing} will depend on this waiting period ($t_w$). Such waiting time dependence is a hallmark of aging systems, and active glasses have been shown to display complex aging behaviors \cite{sollich_mandal_active_aging}. Still, the overall figure's character remains consistent, with poorly annealed glasses exhibiting superior annealing within a given window of time compared to well-annealed ones.

To understand these states' mechanical responses, we selected the states along the top-most (least annealed) curve in Fig. \ref{fig:anealing}(a) and conducted the tensile testing simulations under constant strain rates, the results of which are summarised in Fig. \ref{fig:tensile_tests}. In Fig. \ref{fig:tensile_tests}(b), we see the stress response with the activity turned off after annealing. As expected, we find the annealing effects being reflected in the stress-strain curves as well, with the highest stress peak observed for the $f_0 = 1.9$ case. The $f_0 = 0$ curve shows no overshoot in stress, a signature of the ductile yielding process. 
These stress overshoots are preserved even if the pulling simulations are performed in the presence of the activity, as shown in Fig. \ref{fig:tensile_tests}(c). One crucial difference one observes in the stress-strain curve with and without the presence of activity during the shearing process is the nature of the steady state reached by the system at large strain. Without any active force, the system reaches a unique steady state for all the configurations irrespective of their annealing history, a strong signature of attaining ergodicity after yielding, a well-known feature of passive glasses. In contrast, the steady-state stress ($\sigma_{\infty}$) of the system with activity on is very different. $\sigma_{\infty}$ decreases with increasing activity, although stress overshoot increases with increasing activity. We discuss this in further detail in subsequent paragraphs. A state much above the yielding amplitude shows nearly liquid-like behavior indicating fluidization due to the active forces. Thus, somewhat counterintuitively, we see that introducing self-motility can lead to a system favoring a brittle response over a ductile one below a critical strength of self-motility.

\begin{figure}[htbp]
  \centering
  \includegraphics[width=0.5\textwidth]{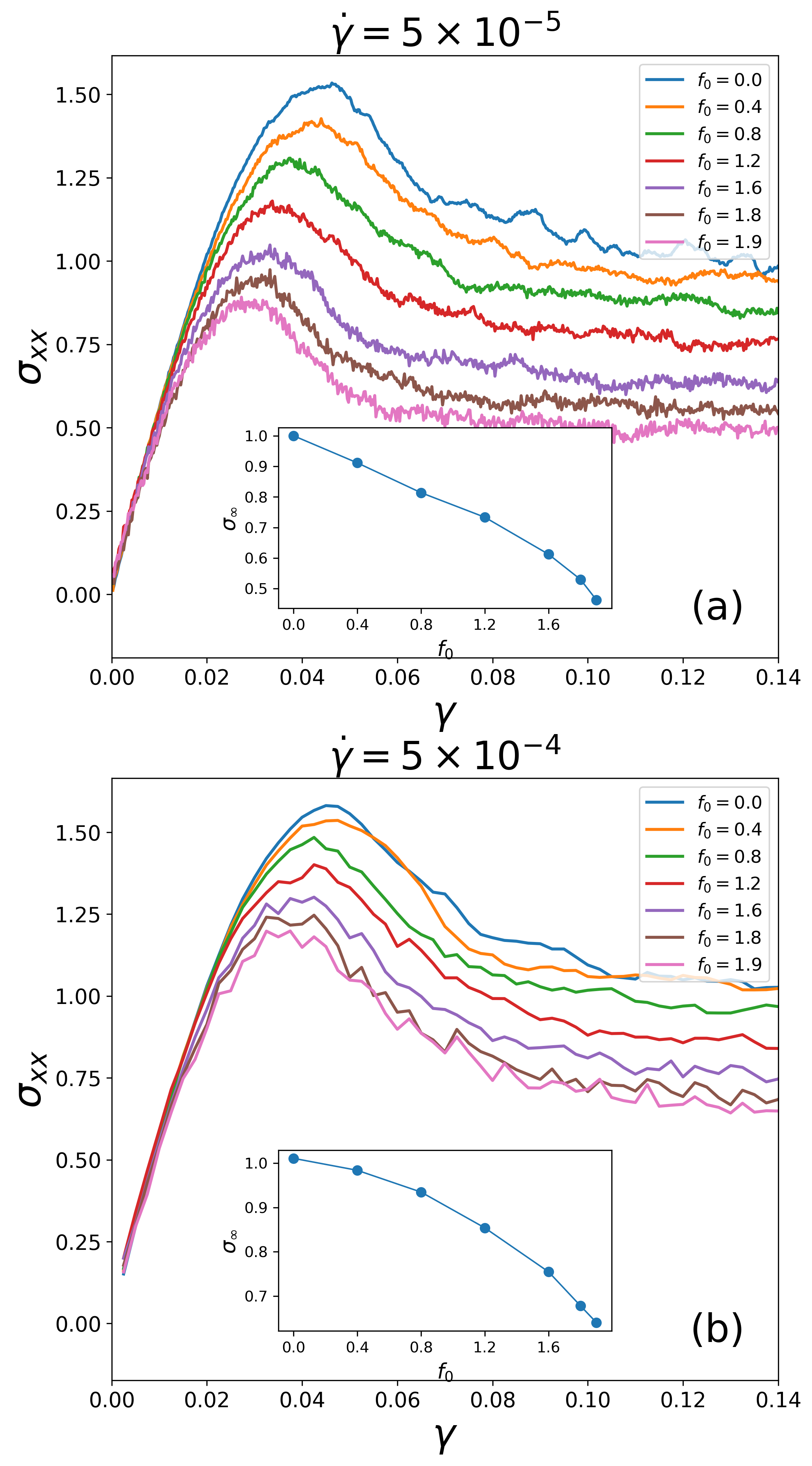}
  \caption{Interplay between intrinsic time scales and imposed time scale (in the form of strain rate) in well-annealed active glasses. A saturation to various pseudo steady states depending on activity showing Herschel-Bulkley-like signature. In the inset, we show the variation of the steady-state stress $\sigma_{\infty}$ with different magnitudes of active forcing, $f_0$. System size of $N = 10,000$ averaged over $16$ ensembles.}
  \label{fig:strain_rate}
\end{figure}

\begin{figure*}[t!]
  \centering
  \includegraphics[width=1.0\textwidth]{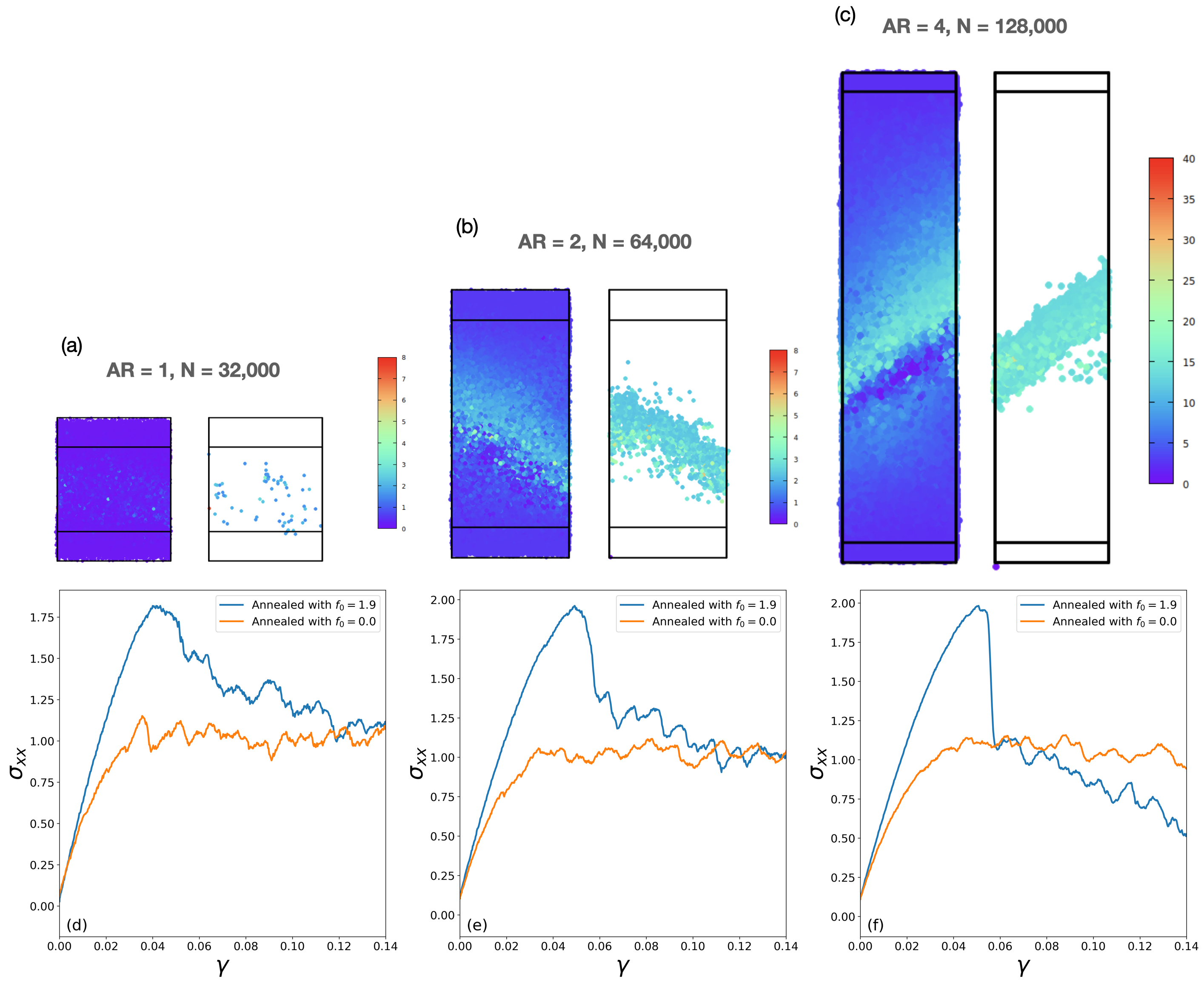}
  \caption{Analyzing the effect of Aspect Ratio (AR) and system size on shear band formation: The formation of shear bands is favored by both a higher aspect ratio and a larger system size. The color bar denotes the magnitude of squared displacements. For an aspect ratio less than 1, a shear band at 45\degree is geometrically impossible to accommodate. For aspect ratios $\leq 1$, hindrance from the walls is significant, and it goes down with larger AR. Thus the same system size considered in (a), but with a higher AR, the system now does show a sharp stress drop; see fig \ref{fig:aspect_ratio_test}. Similarly, the larger system size used in (b), now with decreased AR, ceases to form shear bands. See Fig \ref{fig:aspect_ratio_test2}.}
  \label{fig:shear_bands}
\end{figure*}

\subsection{Effect of strain rate - faster intrinsic time-scales in active glasses}
To study the nature of steady states reached during tensile extension in glasses with active driving present, we looked at the system's response under varying strain rates. 
For this, we took the initial state for testing to be the best-annealed state having per particle inherent structure energy, $E_{IS} = -7.035$, that we achieved via the slowest cooling employed in our simulations. These states were not further subjected to any additional active aging process. This strategy was adopted to separate any potential effects of annealing imposed on the steady states, as we saw in Fig. \ref{fig:tensile_tests}(c). Regardless, as seen from the bottom-most curve in Fig \ref{fig:anealing}(a), an additional annealing treatment for such well-annealed glasses would only have had a minute effect.
In Fig. \ref{fig:strain_rate}, we show the stress-strain curve of the system by changing the strength of the active forces for two different strain rates, $\dot\gamma = 5\times10^{-5}$ and $5\times10^{-4}$. The inset of the respective panels shows how the steady state stress, $\sigma_{\infty}$, decreases with increasing active force, $f_0$.  The qualitative nature of the results can be understood if we consider that with increasing activity, the relaxation time of the system decreases. One can effectively map this behavior to an effective temperature description of a passive system \cite{nandi_Teff}. In a passive system, if one maintains the strain rate but increases the system's bath temperature, then one expects that the steady-state flow stress decreases \cite{JohnsonSamwarPRL, RatulDasgupta} as $\sigma_\infty \sim T^{-2/3}$ across a wide variety of amorphous solids including metallic glasses. Thus it seems that these results can be qualitatively understood using an effective temperature description. In contrast, the annealing effect can not be understood using the same effective temperature description as with increasing the system's effective temperatures; one should observe a reduction in stress overshoot rather than an increase. This suggests that the mechanical response of an active amorphous solid is rather complex, and it is not readily understandable using the simple concept of effective temperature. Thus, further studies along this direction will be needed to explore the full complex rheological behavior of active glasses under various external loading conditions.

As a side note, we want to highlight the effect of activity on the yield strain itself. We observe that introducing activity during tensile testing shifts the yield point towards a lower value, although the magnitude of the stress drop remains fairly unaffected. This is completely different compared to the effect of random pinning on the yield strain as reported in \cite{bhowmik_etal}. Thus active particles can be considered as fluidizing or anti-pinning centers that make the materials less rigid mechanically during deformation. However, it helps to achieve stable mechanical states via annealing when it is allowed to relax.

\subsection{Effect of geometry and larger system size }
To see the catastrophic characteristic of a brittle failure in the stress-strain curve, larger system size and lower inherent state energies are required \cite{ludovicPNAS}. For this, system sizes of $N = 32,000$ , $64,000$, and $128,000$ are considered. To study the effect of geometry and enhanced annealing due to open surfaces, we changed the geometry from a cubical box to a cuboidal rod shape \cite{pinaki_jurgen}. This is done by changing the aspect ratio (defined as $AR = \frac{L_x}{L_y=L_z}$, with x always being the long axis along which the tension is applied) while keeping surface to volume ratio constant. We have considered AR of $1$, $2,$ and $4$ respectively for the three system sizes taken. This gives us an idea of the importance of system size, AR, and absolute surface area during the deformation process. Here, in the original protocol, we introduce an additional time window of $5\times10^6$ MD steps. During this period, we anneal the system in an open geometry while it is still active - that is, after the periodic boundary conditions (PBC) have been removed but before the activity is turned off.  
The importance of understanding the surface dynamics of glassy systems has far-reaching effects on processes like Physical Vapour Deposition (PVD). PVD utilizes the disparity in the mobility of surface particles with orders of magnitude higher mobility than the bulk particle to create ultra-stable glasses~\cite{ediger_etal}. Thus the effects of giving additional mobility are an intriguing avenue to explore.

In Fig. \ref{fig:shear_bands}, we show the appearance of a shear band with increasing aspect ratio and system size. In panel (a), we show that for a system size of $N = 32,000$ particles with $AR = 1$, there is no clear shear band, although, from the overshoot in the stress-strain curve in panel (d), one clearly sees that a well-annealed glass is achieved with $f_0 = 1.9$.
In panel (b), the system size is increased by two folds by increasing $L_x$. Once again, one clearly sees a much sharper stress overshoot in a well-annealed sample prepared with active force $f_0 = 1.9$, as well as a sharper shear band in the middle of the sample at an angle of $45\degree$. With a further increase in system size by increasing $L_x$ by a factor of $2$ again, the stress overshoot becomes nearly discontinuous, as shown in panel (f), with a clear shear band seen in panel (c).

\begin{figure}[htbp]
  \centering
  \includegraphics[width=0.5\textwidth]{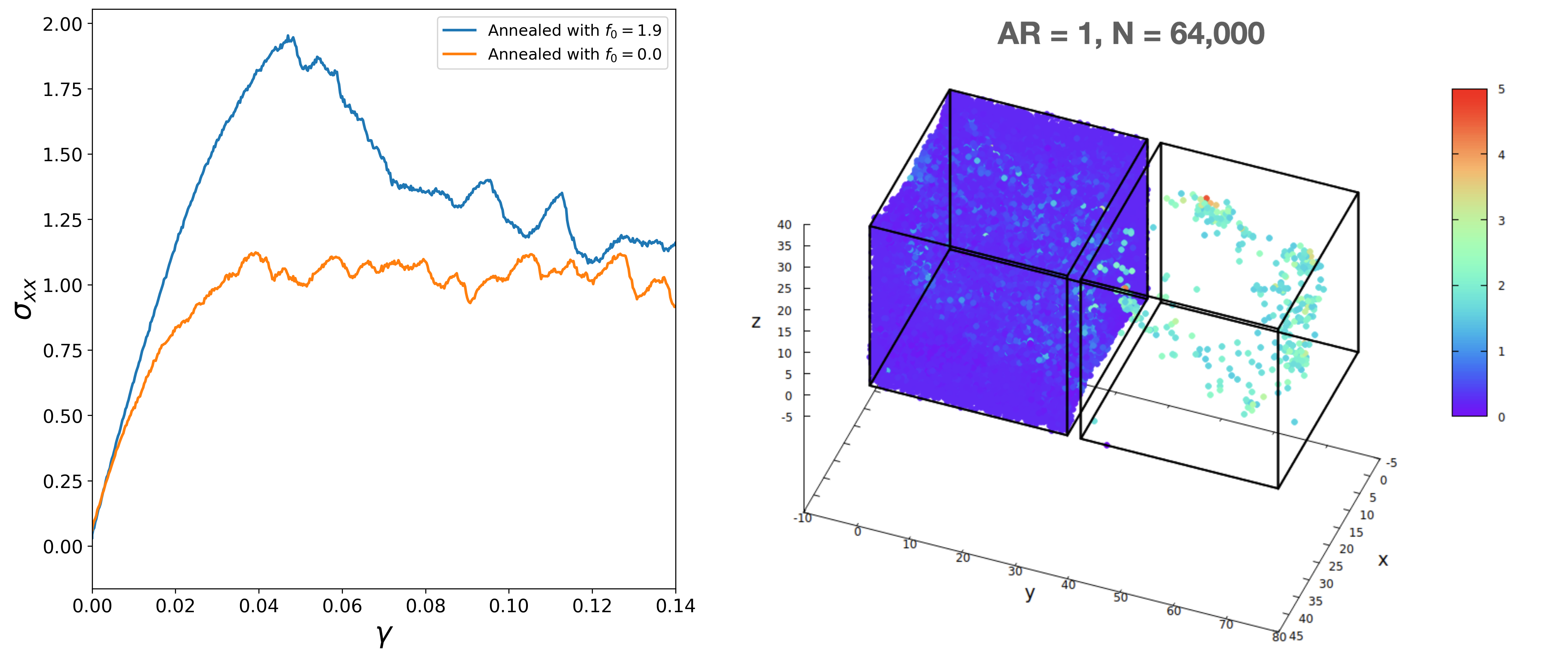}
  \caption{Larger system size but with smaller AR does not form clean shear bands. }
  \label{fig:aspect_ratio_test2}
\end{figure}

\begin{figure}[htbp]
  \centering
  \includegraphics[width=0.5\textwidth]{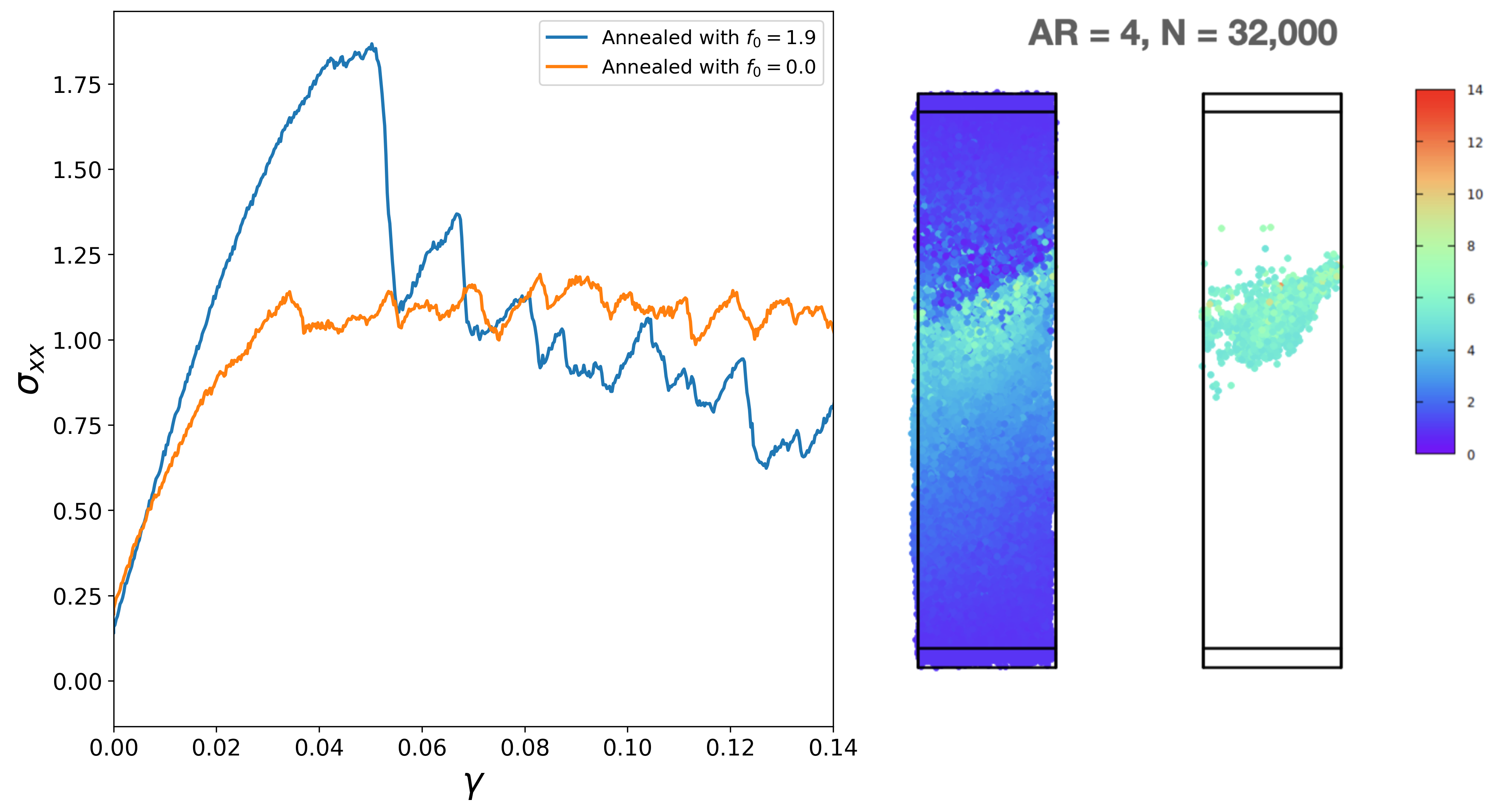}
  \caption{Smaller system size but with higher AR, enough to resolve the shear band. }
  \label{fig:aspect_ratio_test}
\end{figure}

To address the question of whether increasing system size or the increasing aspect ratio plays a dominant role in forming shear bands during failure in well-annealed glasses, we considered the following two scenarios. In the first situation, we take a system size where we clearly observed shear bands for an aspect ratio of 2 (panels b and e in Fig. \ref{fig:shear_bands}) and decreased the aspect ratio to 1 as shown in Fig. \ref{fig:aspect_ratio_test2}. For an aspect ratio of $1$, even for a large system size of $N = 64,000$, the sharpness of the stress drop is decreased significantly compared to when AR was $2$. A lack of a clear shear band in the system accompanies this. This happens because, at this AR, the walls interfere with the formation of the shear band, which happens at a $45\degree$ angle. In the second scenario, we took a system size of $N = 32,000$, which did not show a shear band for an AR of $1$ (panel a and d in Fig. \ref{fig:shear_bands}), but now increased the aspect ratio to $4$, as shown in Fig. \ref{fig:aspect_ratio_test}. In this case, one can see that the stress overshoot becomes nearly discontinuous across the yielding transition, and a clear shear band appears in the sample. This suggests that the geometric shape of the sample, apart from its annealing history, will play a crucial role in determining the eventual failure mode in amorphous solids with open boundaries.

\subsection{Morphology of fractured states}

\begin{figure}[htbp]
  \centering
  \includegraphics[width=0.5\textwidth]{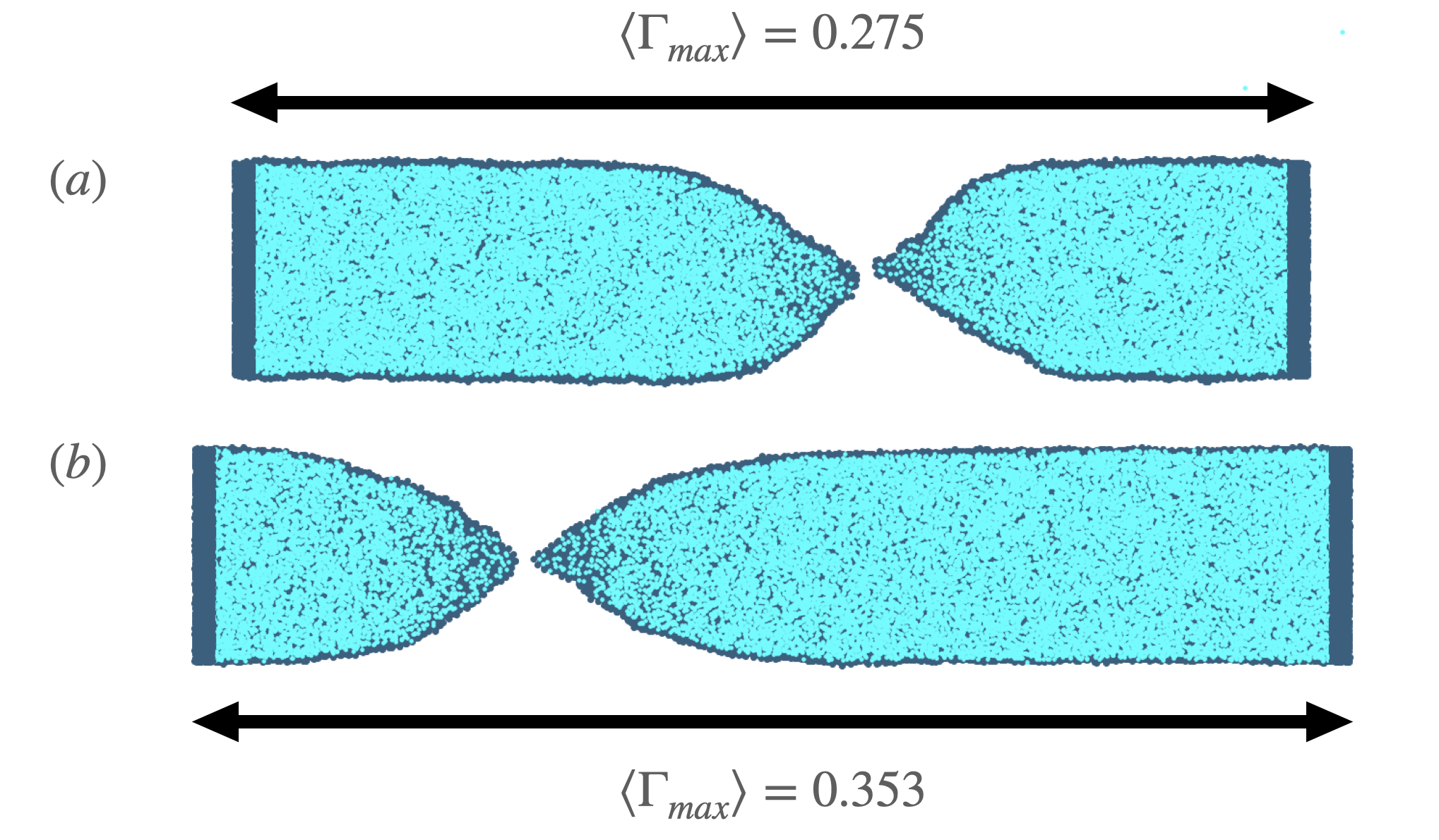}
  \caption{Maximum relative elongation before complete failure. The sample in (a) was treated with activity $f_0 = 1.9$, while the one in (b) was aged without any activity.  The aspect ratio is taken to be $AR = 4$ with system size $N = 128,000$. The strain rate is $\dot{\gamma} = 5\times10^{-5}$. $\Gamma_{max}$ the maximum strain the system can withstand before complete failure. The data is averaged over $3$ ensembles each.}
  \label{fig:gamma_max}
\end{figure}

The percentage elongation of material before fracture, as well as the cross-section of fracture, both provide valuable insights into the failure mechanism. A longer elongation of necking and a thinner cross-section are reminiscent of a ductile failure. Here we show the morphology of the fractured states for two of the samples considered. Sample (a) was treated with activity, whereas sample (b) was just aged passively for the same duration of time. We compute the fractional relative elongation as 
\begin{equation}
\Gamma_{max} = \frac{L_x^f - L_x^0}{L_x^0},   
\end{equation}
where $L_x^0$ is the starting length of the sample in the pulling direction, and $L_x^f$ is the length after the sample detaches into two pieces, as shown in Fig. \ref{fig:gamma_max}. 
We see the average relative elongation in our samples (denoted by $\langle \Gamma_{max} \rangle$ ) after being aged with activity is significantly (nearly 22\% ) smaller than the passively aged case. Thinner necks for the passive case are also observed. This result suggests that activity induces brittleness in the materials. Note that this behavior is completely different than what one expects if the effect of activity can be simply understood as effective temperature. In passive amorphous solids, an increase in temperature leads to more ductile failure. Some of these results are very counter-intuitive, but if one looks at the system in terms of the degree of annealing, then the observation is in agreement with the fact that better-annealed solids will show more brittle type failure behavior.

\section{Conclusion and Perspectives}\label{sec:conclusion}
 In conclusion, we have shown that active particles can effectively anneal glasses to lower energies, just like the mechanical annealing process under oscillatory shear. The similarities observed with the oscillatory shearing suggest that thinking of activity as a form of local shearing rather than just as an effective temperature might provide fruitful insights. By performing extensive tensile testing, we also showed that the annealing effects of activity could be used to change the mode of failure from a ductile to a brittle type. This finding may seem counterintuitive if the activity is simply viewed as a temperature-like phenomenon, considering that higher temperatures generally favor ductile failure. However, it is crucial to recognize the enhanced aging effect in active glasses, which, over time, can transform a ductile active glass into a brittle one simply due to better annealing. Apart from creating better materials, this work can have consequences for developing techniques that are much better at annealing glasses than the usual ones; for instance, a combination of cyclic shear with internal activity might prove to be even better at annealing than any one of those individually.
 Some future research directions could be to see the effect of slightly higher temperatures on the annealing process. Is there an optimal combination of temperature and activity for which a system would anneal the fastest, akin to an optimal combination of shear amplitude and temperature reported in ~\cite{das_etal}? We also plan to explore, in particular, the effect of persistence time in the annealing process. Given the local shear interpretation of active forcing, it seems that higher values of $\tau_p$ would mimic enhanced local random shearing, whereas lower values should have temperature-like effects. Our preliminary findings suggest the same. It would be intriguing to figure out the exact range of $\tau_p$ over which one interpretation of activity should be preferred over the other.
 Could there also be absorbing states if such a study is done in the absence of temperature? This would open the possibilities for encoding memory in such systems, similar to how memory can be encoded in cyclically sheared glasses \cite{memory_srikanth}.
In addition, from an algorithmic perspective, one would like to understand what aspect of activity helps to optimize the search for minima in the complex glassy landscape. Is it the additional search directions available during the persistence time, or is it the stochastic nature of the activity? Answering such questions might help in developing better optimization algorithms in the near future.

\section{Acknowledgment}\label{sec:acknowledgement}
SK acknowledges fruitful discussions with Juergen Horbach during his visit to The Heinrich Heine University, Dusseldorf, Germany. We acknowledge funding by intramural funds at TIFR Hyderabad from the Department of Atomic Energy (DAE) under Project Identification No. RTI 4007. Core Research Grant CRG/2019/005373 from Science and Engineering Research Board (SERB) is acknowledged for generous funding. Most of the computations are done using the HPC clusters bought using CRG/2019/005373 grant and Swarna Jayanti Fellowship, grants DST/SJF/PSA01/2018-19, and SB/SFJ/2019-20/05 of SK. 
\bibliography{references}

\end{document}